\documentclass{article}
\usepackage{arxiv}
\pdfoutput=1

\usepackage[utf8]{inputenc} 
\usepackage[T1]{fontenc}    
\usepackage{hyperref}       
\usepackage{url}            
\usepackage{booktabs}       
\usepackage{amsfonts}       
\usepackage{nicefrac}       
\usepackage{microtype}      
\usepackage{graphicx}
\usepackage{doi}
\usepackage{amsmath}

\title{Advanced models for predicting event occurrence \\
in event-driven clinical trials
accounting for patient dropout, cure and ongoing recruitment}

\date{}

\def\Br{{\rm Br}}

\newtheorem{theorem}{Theorem}[section]

\newtheorem{lemma}[theorem]{Lemma}

\def\a{\alpha}
\def\be{\beta}

\def\la{\lambda}

\def\Si{\Sigma}

\def\Om{\Omega}

\def\th{\theta}

\def\de{\delta}

\def\ka{\kappa}

\def\d{{\rm d}}

\def\E{{\mathbf E}}
\def\Pr{{\mathbf P}}
\def\Va{{\mathbf {Var}}}

\def\cL{{\mathcal L}}

\def\wti{\widetilde}
\def\what{\widehat}

\def\nn{\nonumber}

\def\bee{\begin{equation}\label}
\def\ene{\end{equation}}
\def\beeq{\begin{eqnarray}\label}
\def\eneq{\end{eqnarray}}
\def\beeqn{\begin{eqnarray*}}
\def\eneqn{\end{eqnarray*}}
\def\beth{\begin{theorem}\label}
\def\enth{\end{theorem}}
\def\endd{\end{document}}
\def\beexam{\begin{exmp}\label}
\def\enexam{\end{exmp}}
\def\belem{\begin{lemma}\label}
\def\enlem{\end{lemma}}
\def\becor{\begin{cor}\label}
\def\encor{\end{cor}}
\def\bedf{\begin{df}\label}
\def\endf{\end{df}}
\def\berem{\begin{rem}\label}
\def\enrem{\end{rem}}
\def\bestate{\begin{state}\label}
\def\enstate{\end{state}}
\def\beass{\begin{ass}\label}
\def\enass{\end{ass}}
\def\becons{\begin{cons}\label}
\def\encons{\end{cons}}
\author{
\and
{\large  Vladimir Anisimov\thanks{E-mail: \texttt{Vladimir.Anisimov@amgen.com}} $^1$},
 \
{\large Stephen Gormley$^1$}, \
{\large  Rosalind Baverstock$^1$},\
{\large     Cynthia Kineza$^2$}
}

\renewcommand{\headeright}{}
\renewcommand{\undertitle}{}
\renewcommand{\shorttitle}{Predicting Event Occurrence in Event-Driven Trials}

\hypersetup{
pdftitle={Novel models for predicting clinical events occurrence in event-driven clinical trials
accounting for cure and ongoing recruitment},
pdfauthor={Vladimir Anisimov},
pdfkeywords={Predicting event counts, Patient recruitment,
Event-driven clinical trial, Poisson-gamma recruitment model, Cure model, Dropout},
}

\begin{document}
\maketitle

\vspace{-1.0cm}
\hspace{1.5cm} $^1$Data Science, Center for Design \& Analysis, Amgen, London, UK%

\hspace{1.5cm} $^2$Data Science, Center for Design \& Analysis, Amgen,  Thousand Oaks, CA, US\\[2ex]%

\begin{abstract}

We consider event-driven clinical trials, where the analysis is performed once a pre-determined number
of clinical events has been reached. For example, these events could be progression in oncology
or a stroke in cardiovascular trials.

At the interim stage, one of the main tasks is predicting the number of events over time
and the time to reach specific milestones, where
we need to account for events that may occur not only in patients already recruited and are followed-up but
also in patients yet to be recruited.
Therefore, in such trials we need to model patient recruitment and event counts together.

In the paper we develop a new analytic approach which accounts for the opportunity of patients to be cured,
as well as for them to dropout and be lost to follow-up.

Recruitment is modelled using a Poisson-gamma model developed in previous publications.
When considering the occurrence of events, we assume that the time to the main event and the time to dropout
are independent random variables, and
we have developed a few advanced models with cure using exponential, Weibull and log-normal distributions.
This technique is supported by well developed, tested and documented software.
The results are illustrated using simulation and a real dataset with reference to the developed software.
\end{abstract}


\section{Introduction}

An important aspect of event-driven trials is the operational design at the initial and
interim stages, i.e. predicting the event counts over time and the time to reach specific milestones,
accounting for events that may occur not only in patients already recruited and are followed-up but
also in patients yet to be recruited.
Therefore, in event-driven trials we need to model patient recruitment and event counts together.

There are different techniques for  recruitment modelling described in the literature and one of the main
directions is using mixed Poisson models. This direction has a long history, with several papers devoted
to the use of Poisson processes with fixed recruitment rates to describe the recruitment process
(Carter et al., \cite{carter05}; 
Senn \cite{senn97,senn98}). 
However, in real clinical trials the recruitment rates in different centres vary.
Therefore, to model this variation, Anisimov and Fedorov\cite{anfed07}
introduced a Poisson-gamma model, where the variation in
rates in different centres is modelled using a gamma distribution
(see also Anisimov\cite{an08}).
Some applications to real trials are considered in Anisimov et al.\cite{an-dow-fed07}.
This technique was developed further in several publications
for predicting and interim re-forecasting the recruitment process
under various conditions Anisimov\cite{an11a,an20}.
Other approaches to recruitment modelling primarily deal with global recruitment.
These approaches use different techniques and we refer interested readers to survey papers by
Barnard et al.\cite{barnard10}; Heitjan et al.\cite{heitjan15}) and
Gkioni et al.\cite{gkioni19}, and also to a discussion paper Anisimov\cite{an16b}
on using Poisson models with random parameters with other references therein.

A larger number of clinical trials are event-driven where the number of clinical events
is required to be large enough to allow for reliable statistical
conclusions about the parameters of patient responses.
For such trials, one of the main tasks
is predicting not only the required number of recruited patients,
but also the number of events that may occur
and the time to reach particular milestones.
A useful review of different approaches for event-driven trials is
provided in Heitjan et al.\cite{heitjan15}.
However, for predicting the number of events
over time and the time to stop the trial, the authors of papers cited primarily
use a Monte Carlo simulation technique, e.g.
Bagiella and Heitjan\cite{bagiel2001}.
Therefore, Anisimov\cite{an11b} developed an analytic methodology
for predictive modelling the event counts together
with patient recruitment in ongoing event-driven trials
accounting also for patient dropout.
This methodology is developed further to forecasting
multiple events at start-up and interim stages
under exponential assumptions, Anisimov\cite{an20},
and predicting some operational characteristics
during follow-up times, Anisimov\cite{an16a}.

It is also of interest in event-driven trials that a number of patients
 under treatment will not experience the event within
 their exposure time  (i.e. time from randomisation to a particular milestone)
as the therapy of different diseases is improving.
Therefore, an interesting direction is to investigate the opportunity of cure.
Here we note the paper by Chen\cite{chen16}, but
 he also uses a simulation technique for predicting event timing
 and does not consider the case of patient dropout.

Therefore, in the paper we developed a new analytic approach to this problem
which accounts for patient dropout and also for the opportunity for patients
to be cured with some probability.

We assume that the patient recruitment is modelled using a Poisson-gamma model
developed in Anisimov and Fedorov\cite{anfed07}, Anisimov\cite{an11a}.
We consider non-repeated events and assume that the times to the main event and
to dropout are independent random variables, and there is an opportunity
of cure.
Several new models have been developed using exponential, Weibull and log-normal
distributions.
The focus is on the interim stage, where the parameters of different models
are estimated using maximum likelihood technique.
The predictive distributions of the number of future events for all considered models
are derived in the closed forms, thus, Monte Carlo simulation is not required.

The developed technique and R-tools allow for forecasting  the event counts over time
and also the time to stop the trial with mean and predictive bounds.
The results are illustrated in the paper using Monte Carlo simulation and the real dataset.

The paper is organised as follows:

Section 2, the basic models for the process of event occurrence;
Section 3, predicting event counts for patients at risk; Section 4, predicting event counts accounting
for ongoing recruitment; Section 5, testing of the Weibull model with cure using Monte Carlo simulation;
Section 6, software development and an R-package;
Section 7, implementation to a real clinical trial, and;
Section 8, fitting models to real data.

\section{Modelling the process of event occurrence}
\label{sec2}

Consider a trial at some interim time $t_1$ and assume that there is one type of non-repeated events:  \
the main event of interest $A$, and the patients also can be lost to follow-up (call it dropout).

Then all patients that were recruited in the trial until a given interim time can be divided into three groups:

1) group $A$: patients experienced event $A$.
Denote by $n_A$ the total number of patients in this group
and by $\{x_k \}$ the lengths of follow-up periods
from randomisation date until the event;

2) group $O$: each patient is censored at interim time, thus
the patients have neither experienced an event nor are they lost to follow-up.
Denote by $n_O$ the total number of patients, and by $\{z_i \}$ the lengths of follow-up periods from
randomisation date until interim time;

3) group $L$: patients are lost to follow-up.
Denote by $n_L$ the total number of patients, and by $\{y_j \}$ the lengths of follow-up periods
until censoring by dropout.

Consider now the following cure model describing the process of event occurrence
 for every patient.

Assume that after randomisation, the patient can be either cured with some probability $r$,
or with probability $1-r$ can experience event $A$ after some random time $\tau_A$.
If a patient is cured, then event $A$ cannot occur.

Denote time to dropout $\tau_L$, and if event $A$ doesn't  occur before $\tau_L$,
the patient experiences dropout regardless of whether this patient is cured or not
(in this case event $A$ cannot occur).

Assume that the events for different patients occur independently and
the times $\tau_A$ and $\tau_L$ are also independent random variables with cumulative distribution
functions (CDF) $F_A(x)$ and $F_L(x)$, respectively.
Suppose that
$F_A(x)$, $F_L(x)$ and probability of cure $r$
are the same for all patients, though potentially we can consider
different treatment groups with different parameters.
Assume also that these functions
are continuously differentiable
and denote by $f_A(x)$ and $f_L(x)$ the corresponding probability density functions (pdf).

\subsection{Estimating parameters of the model}\label{MLestimation}

Consider maximum likelihood method.
Denote for convenience, $S_A(x)=1-F_A(x)$ and $S_L(x) = 1- F_L(x)$.

For a patient in group $O$ with exposure time $z_i$,
the probability that event $A$ and dropout will not occur is \
$ S_L(z_i) \Big( r+ (1-r)S_A(z_i)  \Big)  $.

For a patient in group $A$ with exposure time $x_k$,
the probability that event $A$ occurs in a small interval $(x_k,x_k+\d x)$ before dropout is
$(1-r) f_A(x_k) \, S_L(x_i) \d x$.

For a patient in group $L$ with exposure time $y_j$,
the probability that dropout occurs
in a small interval $(y_j,y_j+\d y)$ before event $A$ is
$ f_L(y_i) \Big( r+ (1-r)S_A(y_i) \Big) \d y$.

Given data, the maximum likelihood function has the form
\beeqn
P(F_A,F_L,r) &=&
\prod_{i=1}^{n_O} S_L(z_i) \Big( r+ (1-r)S_A(z_i)  \Big) \\
&\times& \prod_{k=1}^{n_A} (1-r) f_A(x_k) S_L(x_k) \\
&\times&  \prod_{j=1}^{n_L} f_L(y_j) \Big( r+ (1-r)S_A(y_j) \Big)
\eneqn

Correspondingly, the log-likelihood function is
\beeq{LogLikW}
\cL(F_A,F_L,r) &=&  \sum_{i=1}^{n_O} \log(S_L(z_i))
+ \sum_{i=1}^{n_O} \log\Big( r+ (1-r)S_A(z_i)  \Big)  \nn \\
&+& n_A \log(1-r) + \sum_{k=1}^{n_A} \log(f_A(x_k)) + \sum_{k=1}^{n_A} \log(S_L(x_k)) \nn
\\
 &+& \sum_{j=1}^{n_L} \log(f_L(y_j)) + \sum_{j=1}^{n_L} \log\Big( r+ (1-r)S_A(y_j) \Big) \nn
\eneq
For different types of distributions this expression will have a different form.

\subsubsection{Exponential with cure model}

This model assumes that the variables $\tau_A$ and $\tau_L$  are exponentially
distributed with rates $\mu_A$ and  $\mu_L$ respectively.
This is a three parameter model: $(\mu_A, \mu_L, r)$.
The log-likelihood function:
\beeq{LogLik}
\cL(\mu_A,\mu_L,r) &=&  - \mu_A \Si_A -  \mu_L \Si_{1}  + n_A \log(1-r) \nn \\
&+& n_A \log(\mu_A) + n_L \log(\mu_L) \\
&+& \sum_{i=1}^{n_O} \log\Big( r+ (1-r)\exp(-\mu_A z_i )\Big) \nn \\
&+&
\sum_{j=1}^{n_L} \log\Big( r+ (1-r)\exp(-\mu_A y_j )\Big) \nn
\eneq
where $\Si_A = \sum_{k=1}^{n_A} x_k $ and $\Si_1 = \sum_{k=1}^{n_A} x_k + \sum_{j=1}^{n_L} y_j + \sum_{i=1}^{n_O} z_i$.

Consider equating the partial derivatives of the log-likelihood function to zero to find a
relationship between parameters.
Partial derivatives are:
\beeqn
\frac {\partial {\cL(\mu_A,\mu_L,r )}} { \partial r } &=&
- \frac{n_A}{1-r} + \sum_{i=1}^{n_O} \frac{ 1- \exp(-\mu_A z_i )} { r+(1-r)\exp(-\mu_A z_i ) } \\
&+& \sum_{j=1}^{n_L} \frac{ 1- \exp(-\mu_A y_j )} { r+(1-r)\exp(-\mu_A y_j ) } \\
\frac {\partial {\cL(\mu_A,\mu_L )}} { \partial \mu_A } &=&
- \Si_A + n_A/\mu_A \\
&-& (1-r) \sum_{i=1}^{n_O} \frac{z_i\exp(-\mu_A z_i )} { r+(1-r)\exp(-\mu_A z_i )} \\
&-& (1-r) \sum_{j=1}^{n_L} \frac{y_j \exp(-\mu_A y_j )} { r+(1-r)\exp(-\mu_A y_j )} \\
\frac {\partial {\cL(\mu_A,\mu_L )}} { \partial \mu_L } &=&
- \Si_{1} + n_L/\mu_L
\eneqn

Equating the last derivative to zero, we get that
$$
\mu_L = n_L/\Si_1
$$

Substituting into relation (\ref{LogLik}) we get a simpler relation depending only on two variables
$(\mu_A,r)$:
\beeq{LogLik-2}
\cL(\mu_A,r) &=&  - \mu_A \Si_A   + n_A \log(1-r) + n_A \log(\mu_A)  \nn \\
&+& \sum_{i=1}^{n_O} \log\Big( r+ (1-r)\exp(-\mu_A z_i )\Big) \nn \\
&+&
\sum_{j=1}^{n_L} \log\Big( r+ (1-r)\exp(-\mu_A y_j )\Big) \nn \\
&+& n_L (\log(n_L) - \log(\Si_1) - 1) \nn
\eneq

To find the estimators, optimisation is carried out by maximising the log-likelihood function. Initial values are set as
$\mu_A (0) = \frac{n_A}{\Si_1}$ and $r(0)$ taken to be some range of values in $(0, 1)$.
In optimisation, new variables $(\th_1, \th_2)$ are considered:
$$
\mu_A = \exp(\th_1); \qquad  r = \frac{\exp(\th_2)}{1 + \exp(\th_2)}
$$
After optimisation, the variables are transformed back to the original parameters.

\subsubsection{Weibull with cure model}

By definition, the pdf  and CDF
of a Weibull distribution are
$$
f_W(x,\a,b) =  \frac \a {b^\a}  x^{\a-1} e^{-(x/b)^\a}, \, F_W(x,\a,b) = 1- e^{-(x/b)^\a}, \, x > 0
$$
where $(\a,b)$ are shape and scale parameters.
For ease of notation, we use the parametrisation $g=1/b^\a$. Then pdf and CDF have the form
$$
\wti f_W(x,\a,g) = \a g x^{\a-1} e^{-g x^\a},  \
\wti F_W(x,\a, g) = 1- e^{-g x^\a},\, x > 0
$$
Weibull with cure model assumes that the variables $\tau_A$ and $\tau_L$  have Weibull distribution
with parameters ($\a_A, g_A)$ and  $(\a_L, g_L)$ respectively.
This is a five parameter model: $(\a_A, g_A, \a_L, g_L, r)$.
The log-likelihood function:
\beeqn
\cL(\a_A, g_A, \a_L, g_L, r) &=& -g_L \sum_{i=1}^{n_O} z_{i}^{\a_L} +
\sum_{i=1}^{n_O} \log\Big(r + (1 - r)\exp(-g_A z_{i}^{\a_A})\Big) \\
&+& n_A \Big(\log(1 - r) + \log(\a_A) + \log(g_A)\Big) \\
&+& (\a_A - 1) \sum_{k=1}^{n_A} \log(x_k) - g_A \sum_{k=1}^{n_A} x_{k}^{\a_A} - g_L \sum_{k=1}^{n_A} x_{k}^{\a_L} \\
&+& n_L \Big(\log(\a_L) + \log(g_L)\Big) + (\a_L - 1) \sum_{j=1}^{n_L} \log(y_j) \\
&-& g_L \sum_{j=1}^{n_L} y_{j}^{\a_L} + \sum_{j=1}^{n_L}\log\Big(r + (1 - r)\exp(-g_A y_{j}^{\a_A})\Big)
\eneqn
Optimisation is carried out in the same way as for the exponential model, with initial values:
$
\a_A(0) = 1; g_A(0) = n_A/\Si_1;
\a_L(0) = 1; g_L(0) = n_L/\Si_1;
$
$r(0)$ taken to be some range of values in $(0, 1)$.
The new variables $(\th_1,\th_2,\th_3,\th_4,\th_5)$ are:
$$
\a_A = e^{\th_1}; g_A = e^{\th_2};
\a_L = e^{\th_3}; g_L = e^{\th_4};
\ r = \frac {e^{\th_5}} {1+e^{\th_5}}
$$

Similar relations can be written for the combination of the distributions,
e.g. Weibull distribution for time to event $\tau_A$ and exponential distribution for time
to dropout $\tau_L$, and vice versa.

Note that the Weibull model is in some sense a generalisation of the exponential model.
Indeed,  if in particular in the relations above we fix the value $\a_A = 1$, then
we get the combined exponential-Weibull with cure model
(time to event $\tau_A$ has an exponential distribution).
By setting both values, $\a_A = 1$ and $\a_L = 1$, we get the exponential with cure model.

In a similar way the log-likelihood function can be derived also for
a log-normal with cure model.

\section{Predicting event counts for patients at risk }
\label{sec31}

Let us introduce for convenience the time of the occurrence of event $A$, $\nu_A$, so
$\Pr(\nu_A \le z) = (1-r)F_A(z)$.
Note that if $r > 0$, then $\nu_A$ is an improper random variable as $\Pr(\nu_A < +\infty) = 1-r < 1$.

Consider a conditional probability for a patient in group $O$ to experience an event
in the future time interval $[t_1, t_1 + x]$
given that the follow-up period until the interim time $t_1$ is $z$:
\beeq{pAxz}
p_{A}(x,z) &=& \Pr( \nu_A \le z+x, \tau_L > \nu_A \mid \nu_A > z, \tau_L > z ) \nn
\\
 &=& \frac {\Pr( z < \nu_A \le z+x, \tau_L > \nu_A  ) }
{\Pr (\nu_A > z, \tau_L >z )} \\
&=& \frac  {(1-r)\int_{z}^{z+x} f_A(u) S_L(u) \d u }{ S_L(z) \Big( r+ (1-r)S_A(z)  \Big)} \nn
\eneq

For the exponential model $p_{A}(x, z)$ can be calculated in a closed form:
\bee{pAxzExp}
p_{A}(x, z) = \frac{\mu_A}{\mu}\frac{(1 - r)  e^{-\mu_A z} (1 - e^{-\mu x})}{r + (1 - r) e^{-\mu_A z}}
\ene
where $\mu = \mu_A + \mu_L$.

Note that for the exponential model, if $r=0$, $p_{A}(x, 0) = \frac{\mu_A}{\mu} (1 - e^{-\mu x})$,
so this expression does not depend on $z$ and we have a memoryless property. However, for $r >0$,
the memoryless property is lost.

For the Weibull with cure model $p_{A}(x, z)$ has the following form:
\bee{Acurexz}
p_{A}(x,z) = \frac { (1-r)W_2(x,z,\a_A, g_A, \a_L, g_L) }
{ \exp(-g_L z^{\a_L}) \Big( r+ (1-r)\exp(-g_A z^{\a_A}) \Big)}
\ene
where
\bee{pr10}
W_2(x,z,\a_A, g_A, \a_L, g_L) = \a_A g_A \int_z^{z+x} u^{\a_A-1} \exp(-g_A u^{\a_A} -g_L u^{\a_L}) \d u
\ene

To compute this function in applications
we can use a numerical integration.

Similar relations can be written for the combination of the distributions,
and also for a log-normal with cure model.

\subsection{Global prediction}\label{Glob-risk}

Assume now that the recruitment of new patients is already completed,
thus, the events in the future may occur only in patients at risk in group $O$.
Denote by $R_O( t_1, t, \{z_{k}\})$ the total predictive number of events $A$
that may occur in future time interval \ $[t_1, t_1+t]$ for patients in group $O$
where $ \{z_{k}\}$ are the times of exposure.
Let $\Br(p)$ be a Bernoulli random variable, $\Pr(\Br(p)=1) = 1 - \Pr(\Br(p)=0) = p$.

\belem{Lem1}
The process $R_O( t_1, t, \{z_{k}\})$ can be represented in the form:
\bee{riskO}
R_O( t_1, t, (z_{k})) = \sum_{k \in O} \Br(p_A(t,z_k))
\ene
\enlem
where the variables $\Br(p_A(t,z_k))$ are independent and
the probability $p_A(t,z)$ is defined above in Section \ref{sec31} and depends on the type
of the distributions used in the event model.

For a rather large number of patients in group $O$, $(> 20)$, we can apply a normal approximation
for the process $R_O( t_1, t, \{z_{k}\})$
using simple formulae for the mean and the variance:
\beeq{MeanVar}
M(t_1,t) = \sum_{k \in O} p_A(t,z_k),\ 
V^2(t_1,t) = \sum_{k \in O} p_A(t,z_k)(1-p_A(t,z_k))
\eneq
Then $\E[R_O( t_1, t, \{z_{k}\}] = M(t_1,t)$
and
$(1-\de)$-predictive interval at time $t_1+t$ is
$
\Big( M(t_1,t) - z_{1-\de/2}V(t_1,t), M(t_1,t) + z_{1-\de/2}V(t_1,t)) \Big)
$,
where $z_a$ is an $a$-quantile of a standard normal distribution.

For a not so large number of patients, a distribution of $R_O( t_1, t, \{z_{k}\})$
can be calculated numerically as a convolution of the sum of Bernoulli variables.

Let us evaluate the predictive distribution for the time to reach a given target $K$
for the total planned number of events in the study.

Recall that in previous notation $n_A$ denotes the total number of events that occurred prior
to interim time $t_1$ (size of group $A$).
The remaining number of events that are left to achieve
is
$K_R = K - n_A$.

Let $\tau(t_1,K_R)$ be the remaining time  to reach $K_R$ events after the interim time $t_1$.
Then the following relation holds: for any $t > 0$,
\bee{time}
\Pr(\tau(t_1,K_R) \le t) = \Pr(  R_O( t_1, t, \{z_{k}\}) \ge K_R )
\ene
As the distribution of $R_O( t_1, t, \{z_{k}\})$ can be evaluated for any time $t$,
this relation allows us to calculate also the distribution of $\tau(t_1,K_R) $.

Consider the calculation of 
PoS (probability to complete study before a planned time $t_1+T$).
Denote it
as $Q( t_1, T, \{z_{k}\})$.
From (\ref{time}) we get
\bee{PoS}
Q( t_1, T, \{z_{k}\}) = \Pr( R_O( t_1, T, \{z_{k}\}) \ge K_R  )
\ene

If we use a normal approximation for the process $R_O( t_1, T, (z_{k}))$, then
\beeq{PoS2}
Q( t_1, T, \{z_{k}\}) &\approx& \Phi\Big( \frac { M(t_1,T) - K_R}{V(t_1,T)} \Big)
\eneq
where $\Phi(x)$ is the CDF of a standard normal distribution.

\section{Predicting event counts accounting for ongoing recruitment }

Consider now the situation when at the interim time the planned number of patients
to be recruited is not reached yet, that means, the recruitment is still ongoing.
In this case we need also to predict the future recruitment
and how many events may occur for patients to be recruited in the future.

\subsection{Modelling and predicting patient recruitment }\label{recruit}

Assume that patients arrive at clinical centres according to Poisson processes
with some rates $\la_i$. To model the variation in the rates among different
centres we assume that $\la_i$ are jointly independent gamma distributed random variables
with parameters $(\a,\be)$ (shape and rate) and pdf
\begin{equation}\label{e00}
f(x,\alpha,\beta) =
\frac {e^{- \beta x} \beta^{\alpha} x ^{\alpha-1} }{ \Gamma(\alpha)},\ x >0,
\end{equation}
where
$ \Gamma(\alpha)$ is a gamma function.

This model is called a Poisson-gamma (PG) recruitment model and was developed in
Anisimov \& Fedorov\cite{anfed07} and further extended in
Anisimov\cite{an08,an11a,an20}.

Denote by $\Pi_a(t)$ a standard Poisson process with rate $a$
and by $\Pi(a)$ a random variable which has a Poisson distribution with parameter $a$.
Then a mixed Poisson process $\Pi_{\lambda}(t)$ where the
rate $\lambda$ is gamma distributed with parameters $(\a,\be)$ is a
PG process (Bernardo and Smith\cite{bernardo04})
with parameters
$(t,\alpha,\beta)$:
\bee{PG1}
\Pr(\Pi_{\lambda}(t)  =  k)
= \frac{\Gamma(\alpha + k)}{k!\ \Gamma(\alpha)}\ \frac{t^{k}\beta^{\alpha}}
{\ {(\beta + t)}^{\alpha + k}}\ ,\ k = 0,1,2,..
\ene

Note that for a mixed Poisson process with random rate $\la$,
\bee{MeanVar-2}
\E [\Pi_{\la}(t)] = \E[\la] t; \,
\Va [\Pi_{\la}(t)] = \E[\la] t + \Va[\la] t^2
\ene

Assume now that some centre is active only in time interval $[u,b]$.
Denote by $d(t, u, b)$ the duration of recruitment window
(duration of active recruitment)
in a centre up to time $t$:
\bee{dtab}
	d(t, u, b) =
	\begin{cases}
		0 & t \leq u \\
		t - u & u < t \leq b \\
		b - u & t > b
	\end{cases}
\ene

Assume that the recruitment rate in this centre is $\la$ which is gamma distributed with some parameters.
Then the recruitment process in this centre for any $t > 0$ can be represented as a PG process
with a cumulative rate $\la d(t,u,b)$. That means, the number of patients recruited in interval
$[0,t]$ has a mixed Poisson distribution with the rate $\la d(t,u,b)$.

Consider now predicting the remaining recruitment at some interim time $t_1$.
Assume for simplicity that all centres are active and in every centre $i$
the following data are available: \
$(v_i,k_i)$ - the duration of active recruitment (recruitment  window) and the number of patients recruited.

In
Anisimov and Fedorov\cite{anfed07} (see also Anisimov\cite{an11a}) it was developed a
maximum likelihood technique for estimating parameters $(\a,\be)$ of a PG model
assuming that in all active centres the rates have a gamma distribution with the same parameters.
In \cite{an11a}, the Bayesian technique was also developed
for predicting future recruitment using the property
that the posterior rate in a centre $i$, $\widetilde{\lambda_{i}}$, which is adjusted to
the data in this centre, also has a gamma distribution
with parameters $(\a+k_i,\be+v_i)$.

Consider now a given interim time $t_1$. Let $i$ be some active centre.
Denote by $(\a,\be)$ the parameters of a PG model estimated
using data in all active centres as noted above.
Then the future recruitment process in centre $i$ can be modelled as a PG process
with posterior recruitment rate $\widetilde{\lambda_{i}}$.
Assume that the recruitment in this centre can be closed due to some operational reasons
at some time $t_1 + b_i$.
Then for any $t>0$ the recruitment process in centre $i$ in time interval $[t_1,t_1+t]$
can be represented as a PG process with a cumulative rate
$\wti \la_i d(t,0,b_i)$.

Assume now that $j$ is some new centre that is planned to be initiated at time $t_1+u_j$
and let $b_j$ be the closing date of recruitment in this centre.
Denote by $\la_j$ the recruitment rate in this centre.
Note that the rates in the new centres can be provided by clinical teams
using expert estimates or evaluated using historical data from similar trials.

Then centre $j$ will be active only in time interval $[t_1+u_j,t_1+b_j]$.
Thus, for any $t>0$, the recruitment process
in time interval $[t_1,t_1+t]$
can be represented as a PG process with a cumulative rate
$\la_j d(t,u_j,b_j)$.

Consider the prediction of the remaining global recruitment.

Denote by $I_{Active}$ a set of active centres with posterior rates $\widetilde{\lambda_{i}}$.
Assume also that it can be some set $I_{New}$ of new centres that are planned to be
initiated after interim time $t_1$ at times  $t_1+u_j,\, j \in I_{New}$.
Denote by $\la_j$ the rates in the new centres.
Then the predictive total number of patients $n(t_1,t_1+t)$ to be recruited in the time interval
$[t_1,t_1+t]$ can be represented as
\bee{PredRecr}
n(t_1,t_1+t) = \sum_{i \in I_{Active}} \Pi(\wti \la_i d(t,0,b_i))
+ \sum_{j \in I_{New}} \Pi(\la_j d(t,u_j,b_j))
\ene

This means, $n(t_1,t_1+t)$  has a mixed Poisson distribution with a cumulative rate
\bee{PredRecr-2}
\Om(t_1,t_1+t) = \sum_{i \in I_{Active}} \wti \la_i d(t,0,b_i)
+ \sum_{j \in I_{New}} \la_j d(t,u_j,b_j)
\ene

For a rather large number of centres, the predictive bounds for $n(t_1,t_1+t)$
can be evaluated using a normal approximation, as the mean and the variance
of $n(t_1,t_1+t)$ can be easily calculated using the property (\ref{MeanVar-2})
and relations \
$
\E [\tilde{\lambda_{i}} ] = (\alpha + k_{i})/(\beta + v_{i}); $
$\Va [\tilde{\lambda_{i}} ] = (\alpha + k_{i})/(\beta + v_{i})^2
$.
In particular, the mean predicted time to reach a required remaining number of patients
$n_R$ can be numerically calculated as the point when the line $\E[n(t_1,t_1+t)]$
hits level $n_R$.

Note that for a not so large number of centres, for predicting $n(t_1,t_1+t)$ one can use
a PG approximation developed in Anisimov\cite{an20}, Anisimov and Austin\cite{{an-austin20}}.

\subsection{Predicting event counts}

Consider now predicting event counts accounting for ongoing recruitment.

Denote by
\(\ka_{A}\) the time it takes until event \(A\) occurs first (before dropout), and let
$p_{A}( x ) = \Pr(\ka_{A} \leq x)$, $x > 0$, be its CDF.

For cure model with dropout defined in Section \ref{sec2}, in previous notation,
\bee{pAt}
p_{A}( x ) = \Pr( \nu_A \le x, \nu_A < \tau_L ) = (1-r)\int_{0}^{x} f_A(u) S_L(u) \d u
\ene

In particular, for the exponential with cure model, using notation $\mu = \mu_A + \mu_L$,
\bee{ExpMod}
p_{A,E}( x ) = (1-r) \frac{\mu_A}{\mu} (1-e^{ -\mu x})
\ene

For Weibull model, using parametrisation  $\bar \theta = (\a_A, g_A, \a_L, g_L)$,
\bee{pr20}
p_{A,W}(x) = (1-r) W_1(x,\bar \theta)
\ene
where
\bee{pr21}
W_1(x,\bar \theta) = \a_A g_A \int_0^x y^{\a_A-1} \exp(-g_A y^{\a_A} -g_L y^{\a_L}) \d y
\ene

Consider
now one clinical centre. Assume that patients
arrive according to a mixed Poisson process
with possibly random rate $\la$.
Assume also that the centre is active only in a fixed time
interval $[a,b]$.
In Anisimov\cite{an11b,an20}
the following result is proved.

\belem{Lem2}
The predicted number of events \(A\) in interval
$[0,t]$ that occur in
the newly recruited patients in this centre
has a mixed Poisson distribution with rate $\la q_A(t,a,b)$, where 
\bee{8.29}
q_A(t,a,b) = \int_{a}^{min(t,b)} p_{A}( t - u)\d u
\ene
\enlem

For the exponential model, the function $q_A(t,a,b) $ can be easy calculated.
Consider the duration of recruitment window $d(t, a, b)$ in a centre at time $t$
defined in (\ref{dtab}).
Then, using parameters $(r,\mu_A,\mu_L)$,
\bee{ExpMod-2}
q_{A,E}(t,a,b)
= (1-r) \frac{\mu_A}{\mu} \Big( d( t,a,b) - \frac{1}{\mu}e^{- \mu ( t - a)}
( e^{\mu d( t,a,b)} - 1) \Big)
\ene

For Weibull Model
with parameters $(\a_A, g_A, \a_L, g_L, r)$,
\bee{WeiMod}
q_{A,W}(t,a,b)
= (1-r)  \int_{a}^{min(t,b)} W_1(t-u,\bar \theta) \d u
\ene
where $W_1(t,\bar \theta)$ is defined in (\ref{pr21}).
This function  can be  numerically calculated.

Similar relations in the integral form
can be written for the combination of the distributions,
and also for a log-normal with cure model.

These results
form the basis for creating predictions of the
event counts in any active centre and globally.

\subsection{Global forecasting event counts at interim stage }

Consider now forecasting the total number of events at some interim time $t_1$.

Denote the times of initiation of new centres (if any) by $\{ u_i \}$  and the times of closure
for all centres by $\{ b_i \}$.
In general it is assumed that centres will be closed
for recruitment at the time when recruitment hits the recruitment target.
Thus,
in applications, we usually assume that $b_i \equiv \what T_{Pred}$ where
$ \what T_{Pred}$ is the predicted mean remaining time to reach the recruitment target.


\beth{Th1}
The predictive total number of new events $A$, $k(t_1,t,A )$, that may
occur in future time interval $[t_{1},t_{1} + t]$,
can be represented as a convolution of two independent
random variables:
\bee{eqn1}
k(t_1,t,A ) = \Pi( \Sigma( t,A) )+ R_O( t_1, t, \{z_{k}\})
\ene
where according to (\ref{PredRecr-2}),
\bee{eqn2}
\Sigma\left( t,A  \right) =
\sum_{i \in I_{active}} \tilde{\lambda_{i}} q_{A}( t,0,b_i ) +
\sum_{i \in I_{new}}
\lambda_{i} q_{A}( t,u_{i},b_i ),
\ene
and $R_O( t_1, t, \{z_{k}\})$ is the predictive number of events $A$
in group $O$ defined in (\ref{riskO}), Section \ref{Glob-risk}.

Here
the function $q_{A}(t,a,b) $ is defined by (\ref{8.29}) (for exponential and Weibull
models we have the expressions (\ref{ExpMod-2}) and (\ref{WeiMod}), respectively).
The first sum in (\ref{eqn2}) is taken across all active centres and
$\tilde{\lambda_{i}}$ are the posterior rates defined in Section \ref{recruit}.
and the second sum is taken across new centres.

Correspondingly, the probability to complete trial in time is
\bee{eqn3}
\Pr \Big(k( t_1, T_{R},A) \geq \nu_{R}( A ) \Big)
\ene
where $T_{R}$ is the planned remaining time to complete the trial and
$\nu_{R}( A )$ is the remaining number of events left to achieve.
\enth

The proof follows from results of Lemmas \ref{Lem1}, \ref{Lem2}
and Section \ref{recruit}.

Note that the mean and the variance of the process $\Pi( \Sigma( t,A) )$
can be calculated explicitly in terms of functions $q_A(\cdot)$ and parameters of the rates.

As typically in real trials the number of centres is rather large, to create predictive
bounds for $k(t_1,t,A )$ one can use a normal approximation.
This technique is realised in R package (\textit{EventPrediction}), see Section \ref{package}.

\section{Monte Carlo simulation}
\begin{figure}[ht]
\begin{center}
\includegraphics[width=11.0cm,height=6cm]{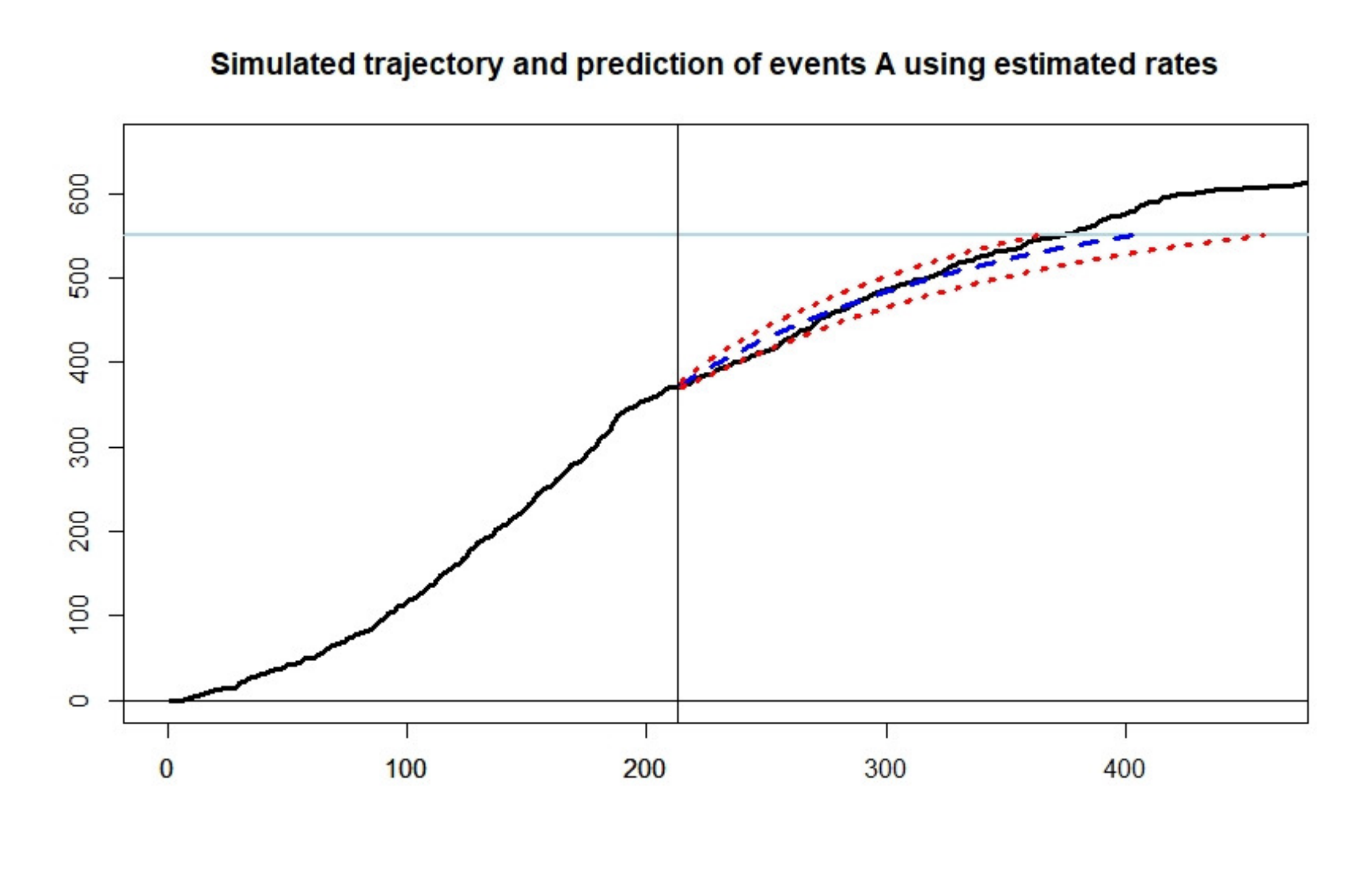}
\caption{Plot of number of events against time (in days), following the timeline of a simulated trial.
The simulated trajectory of events A is marked by the black solid line, the initial parameters: $a_A = 0.8$,
$b_A = 182$, $a_L = 0.6$, $b_L = 2611$ and $r = 0.2$. An interim analysis was taken at 7 months,
the estimated parameters: $a_A = 0.842$, $b_A = 145$, $a_L = 0.641$, $b_L = 2697$ and $r = 0.276$.
Predictions on future event counts were created using the estimated parameters; the mean trajectory
is shown by the blue dashed line, the 90\% confidence bounds by the red dotted lines.
}
\label{fig:aA<1}
\end{center}
\end{figure}

\begin{figure}[ht]
\begin{center}
\includegraphics[width=11.0cm,height=5cm]{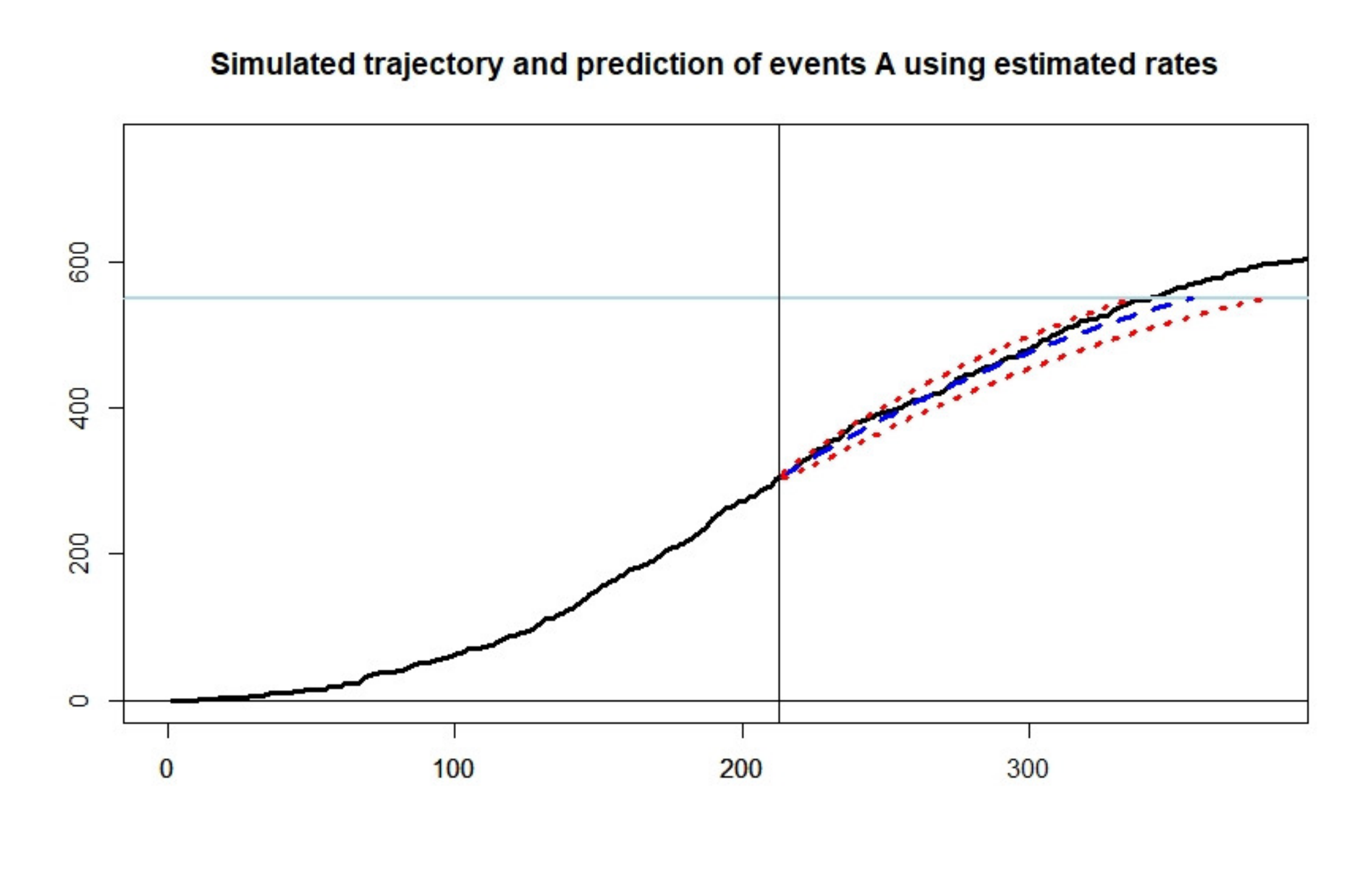}
\caption{Plot of number of events against time (in days), following the timeline of a simulated trial.
The simulated trajectory of events A is marked by the black solid line, the initial parameters: $a_A = 1.2$,
$b_A = 213$, $a_L = 1.4$, $b_L = 3701$ and $r = 0.2$. An interim analysis was taken at 7 months,
the estimated parameters: $a_A = 1.265$, $b_A = 175$, $a_L = 1.406$, $b_L = 3916$ and $r = 0.305$.
Predictions on future event counts were created using the estimated parameters; the mean trajectory
is shown by the blue dashed line, the 90\% confidence bounds by the red dotted lines.}
\label{fig:aA>1}
\end{center}
\end{figure}

Monte Carlo simulation was used to test each model's performance.
We considered 1000 patients assuming uniform distribution of centre initiation
over 6 months, and took the target number of events  550.
At a specified cut-off date the model parameters were estimated
using maximum likelihood estimation, see Section \ref{MLestimation}.
Using these estimators, predictions of the future occurrence of events were created.
For the Weibull model two different cases for the initial parameters were considered,
$a_A < 1$ and $a_A > 1$, see Fig~\ref{fig:aA<1} and Fig ~\ref{fig:aA>1} respectively.

In both cases, the model successfully predicts the trajectory of the number
of events $A$
with the real trajectory falling within the 90\% predicted bounds.
Furthermore, the parameters estimated at the cut-off time using
maximum likelihood technique are close to the initial parameters showing an appropriate estimation.

\section{Software development}\label{package}

In order to expose the event and recruitment prediction models (as detailed in the previous sections)
to a large number of key stakeholders, an R package (\textit{EventPrediction}) has been developed,
tested and deployed to a centralised R server.  The \textit{EventPrediction} package allows
a user to easily pass the data required (i.e. subject event data, centre level data and configuration)
and return back key parameter estimates and predictions with bounds, for both events and recruitment.

\subsection{R package design}

R was chosen over other programming languages, and an R Package was developed over
standalone R scripts,
for a number of reasons, including:  R has an easy to use and to setup testing framework;
R ships with easy to use code coverage tools; R has a Comprehensive R Archive Network ("CRAN")
set of packages that are easily accessible; R seamlessly integrates with GitLab
(and other source control software); R allows for an Object Oriented ("OO") approach
(i.e. S3, S4 and R6); and also, primarily, it is simply straightforward to develop, test,
document and centrally deploy an R Package for key stakeholders
to use.

R's S3 lightweight OO solution was a key design feature of the \textit{EventPrediction}
package, as using such an OO approach yields four main benefits: first, S3's simple
to use OO benefit of polymorphism (aka in R as method dispatch); secondly, S3 gives the
OO benefit of inheritance; thirdly, S3 is ubiquitously used by R contributors, easy to use
and for others to comment; and fourthly, S3 is in accordance with the functional programming paradigm,
when an object is passed into an S3 function it is not going to change (unlike full OO approaches like R6).

\subsubsection{Good software engineering principles}
Another major benefit of developing an R Package (and utilising R's OO approach) is
to ensure adherence to good Software Engineering principles, with code that is at a minimum: reliable;
easy to use; efficient; well tested, with tests traceable to requirements and/or design;
well documented; and (importantly) easy to maintain.  The \textit{EventPrediction} package conforms
to each of these key programming elements, not only because these are simply good
Software Engineering practices, but also as the biotechnology sector is highly regulated
and there is a requirement to document a number of Software Development Life Cycle ("SDLC")  tasks in accordance
with departmental, company and regulatory policies.

\subsubsection{R package SDLC and platform architecture}
Before the design, development and/or testing of any code was initiated, two further key platform
architectural design decisions were made: first, GitLab was used for source control, continuous
integration, documentation, vignettes, readme files and also as part of the full deployment process;
and secondly, R Studio Server Pro was used for development and testing of code,
a Docker Image with a physical R server on AWS.

\subsubsection{Further R package design: function layers}

With a large number of complex R scripts and source papers another design choice (primarily,
to make the code easier to use and easier to maintain) was grouping the code into four layers
using R's S3 OO approach.
The four programming layers are as follows:

{\em Layer One: Highest Level: Main Exposed Application Programming Interface (API).}

This level is exposed to the user and contains: S3 Classes (functions) that allow
instantiation of the objects that contain the input data required and configuration;
functions to predict events and recruitment; plotting and printing functionality;
and key getter functions.

{\em Layer Two: Second Level Functions.}

This level is not exposed to the user and is simply used to dispatch to the third level
functions based on the S3 configuration objects instantiated in Layer One.

{\em Layer Three: Third Level Functions.}

This level is not exposed to the user and contains the main set of controller code
and does all of the hard work of the package.

{\em Layer Four: Lowest Level Functions.}

This level is not exposed to the user and contains a large number of complex
R scripts/algorithms that have been developed and tested using
Monte Carlo Simulation, as detailed in the previous section.

\subsection{R package input data required}
The following set of input data is required by the \textit{EventPrediction} package
to predict events and recruitment (if recruitment is ongoing), with each set of data instantiated
using R's S3 approach (as detailed in the previous sections).

\subsubsection{Event data}
\begin{center}
\begin{tabular}{c c c c}
analysis\_time\_days & \quad  censor\_flag & \quad drop\_out\_flag & \quad randomisation\_date \\
28 & 0 & 0 & {\small YYYY-MM-DD} \\
33 & 0 & 0 & {\small YYYY-MM-DD} \\
87 & 1 & 0 & {\small YYYY-MM-DD} \\
42 & 0 & 0 & {\small YYYY-MM-DD} \\
77 & 1 & 1 & {\small YYYY-MM-DD}
\end{tabular}
\end{center}
This data is in accordance with how the key stakeholders produce their data,
it is transformed into the values as described in the previous sections, such that:
\begin{itemize}
\item analysis\_time\_days is the number of days from randomisation to either
the event date ${T_A}$ or censoring date (i.e. the dropout date ${T_L}$ for
subjects that have dropped out or the date used to censor at the cut off
if a subject has not dropped out).
\item censor\_flag == 0  represents group A, a subject experienced event A.
\item censor\_flag == 1 \& drop\_out\_flag == 0 represents group O, a subject
did not experience an event nor dropout.
\item drop\_out\_flag == 1 represents group L, a subject dropped out
before the interim time.
\end{itemize}

\subsubsection{Site recruitment data}
\begin{center}
\begin{tabular}{c c c}
study\_centre\_id & \quad centre\_actual\_enrol &\quad  centre\_recruitment\_window\_days \\
xx001&0&140 \\
xx002&1&224 \\
xx003&2&238 \\
xx004&1&221 \\
xx005&0&201
\end{tabular}
\end{center}

\begin{itemize}
\item centre\_actual\_enrol represents the number of subjects
recruited at the unique centre ID.
\item centre\_recruitment\_window\_days represents the actual duration of recruitment
at the unique centre ID (not including any screening period).
The centre is active only during this interval $[a,b]$.
\end{itemize}

\subsubsection {New Sites}
A vector of days for new centres to be initiated \{$u_i$\}, e.g. $c(3, 5, 5, 10, 10, 11, 12, 20)$.

\subsubsection{Configuration}
The following key pieces of information are accepted by the \textit{EventPrediction}
package
(with appropriate defaults) which are used to select the appropriate algorithms
and to provide key modelling values:
\begin{itemize}
\item distributions\_to\_use: A list detailing the distributions to model the dropouts and events:
       e.g. list(events = "Exponential", drop\_outs = "Exponential")
\item target\_number\_of\_events: The target number of events for the analysis to be predicted
\item sample\_size: The number of patients planned to recruit
\item confidence\_level: The confidence probability for the upper and lower bounds
\end{itemize}

\section{R package and implementation in a clinical trial}


\subsection{Introduction}

In order to help the key stakeholders with the operational planning of a clinical trial
and to test the quality of the prediction, the \textit{EventPrediction} package was used
on several historic studies.
The following is one such case study in a historical oncology clinical trial,
using the data at a given interim time when recruitment had not completed.
The task was to predict the future recruitment and event counts with bounds and compare
the results with the real trajectory of the recruitment and the events that have already occurred in the past.

The event and centre data was provided in accordance with the package API's
(as detailed in the previous section) along with a target number of events of 250
and patients sample size of 405.
At the interim cut-off time the data
for the study had the following recruitment and event status:
\begin{itemize}
\item 152 Events (i.e. censor\_flag == 1)
\item 155 At Risk (i.e. censor\_flag == 1 and drop\_out\_flag == 0)
\item 13 Drop Outs( i.e. drop\_out\_flag == 1)
\item
85 patients left to recruit.
\end{itemize}

\subsubsection{Key predictions}
Given the above input and implementing the developed model,
the \textit{EventPrediction} package predicted:

\textit{Recruitment:} Predicted number of days until target number of patients is reached
with 90\% bounds, (mean, lower bound, upper bound): 151, 120, 191.

The estimated parameters of a PG model are:
$\a = 4.8577, \be = 516.13$, and the prediction is constructed according to
(\ref{PredRecr}) where it was used some schedule of closing centres.

\textit{Events:} Predicted number of days until target number of events
reached, with 90\% bounds, (mean, lower bound, upper bound):\\
exponential model: 227, 181, 322 \\
Weibull model: 241, 188, 423

\begin{figure}[ht]
\begin{center}
\includegraphics[width=11.0cm,height=5cm]{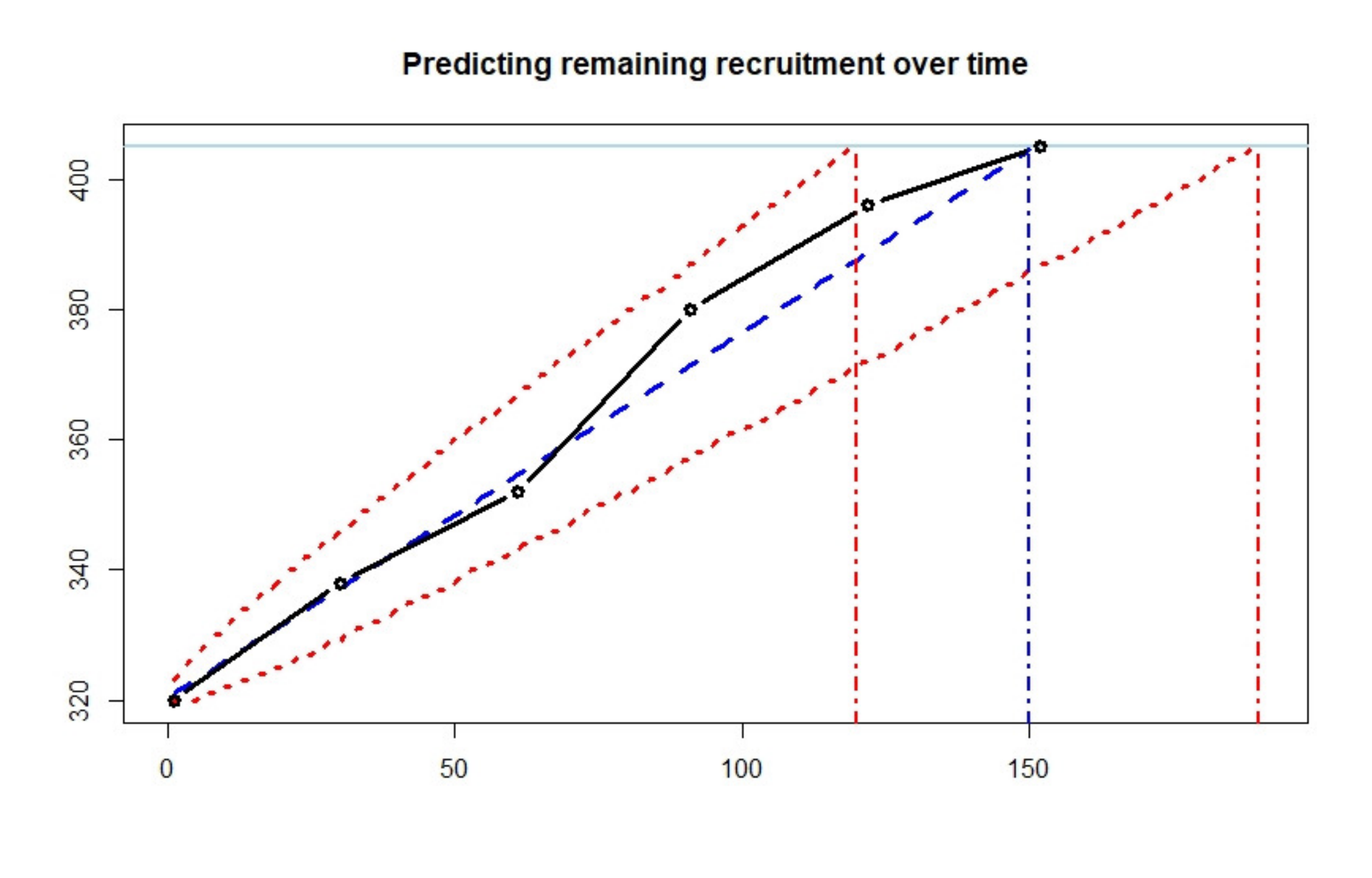}
\caption{Prediction of the remaining recruitment against time from cut-off (in days).
Real trajectory of patient recruitment is shown by black line.
Mean prediction and 90\% bounds are shown by the blue dashed and red dotted lines.}
\label{Fig-4}
\end{center}
\end{figure}

\subsubsection{Plots and parameter estimates}

Further, the \textit{EventPrediction} package produced the following three key plots,
along with key parameter estimates, for the key stakeholders to consume: \\
1) prediction of the remaining recruitment \\
2) prediction of the remaining number of events using exponential model \\
3) prediction of the remaining number of events using Weibull model

\begin{figure}[ht]
\begin{center}
\includegraphics[width=11.0cm,height=6cm]{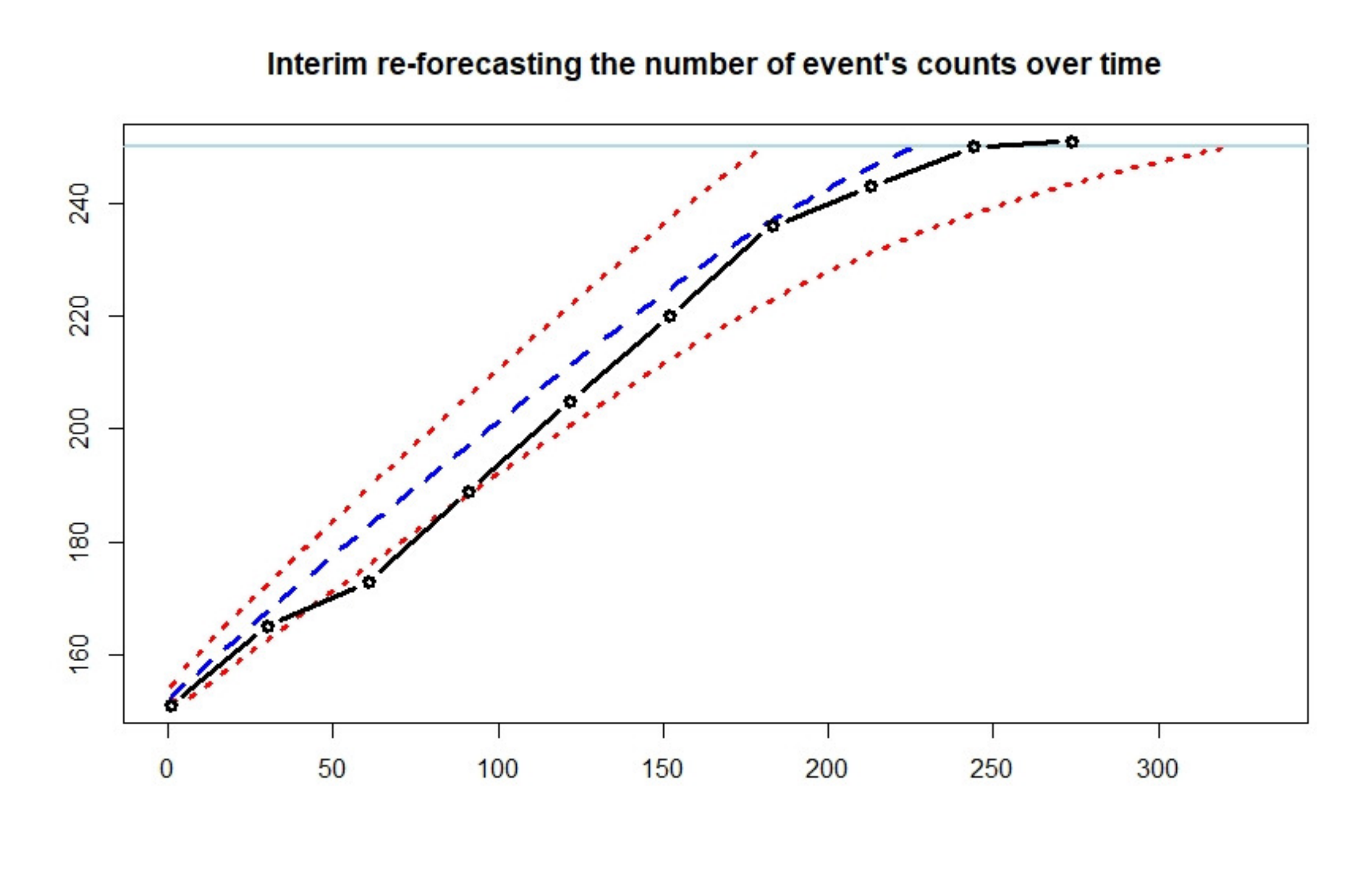}
\caption{Prediction of the remaining number of events against time from cut-off (in days).
Real trajectory of events is shown by black line.
Exponential model, mean prediction and bounds depicted by the blue dashed and red dotted lines.}
\label{Fig-5}
\end{center}
\end{figure}

\begin{figure}[ht]
\begin{center}
\includegraphics[width=11.0cm,height=6cm]{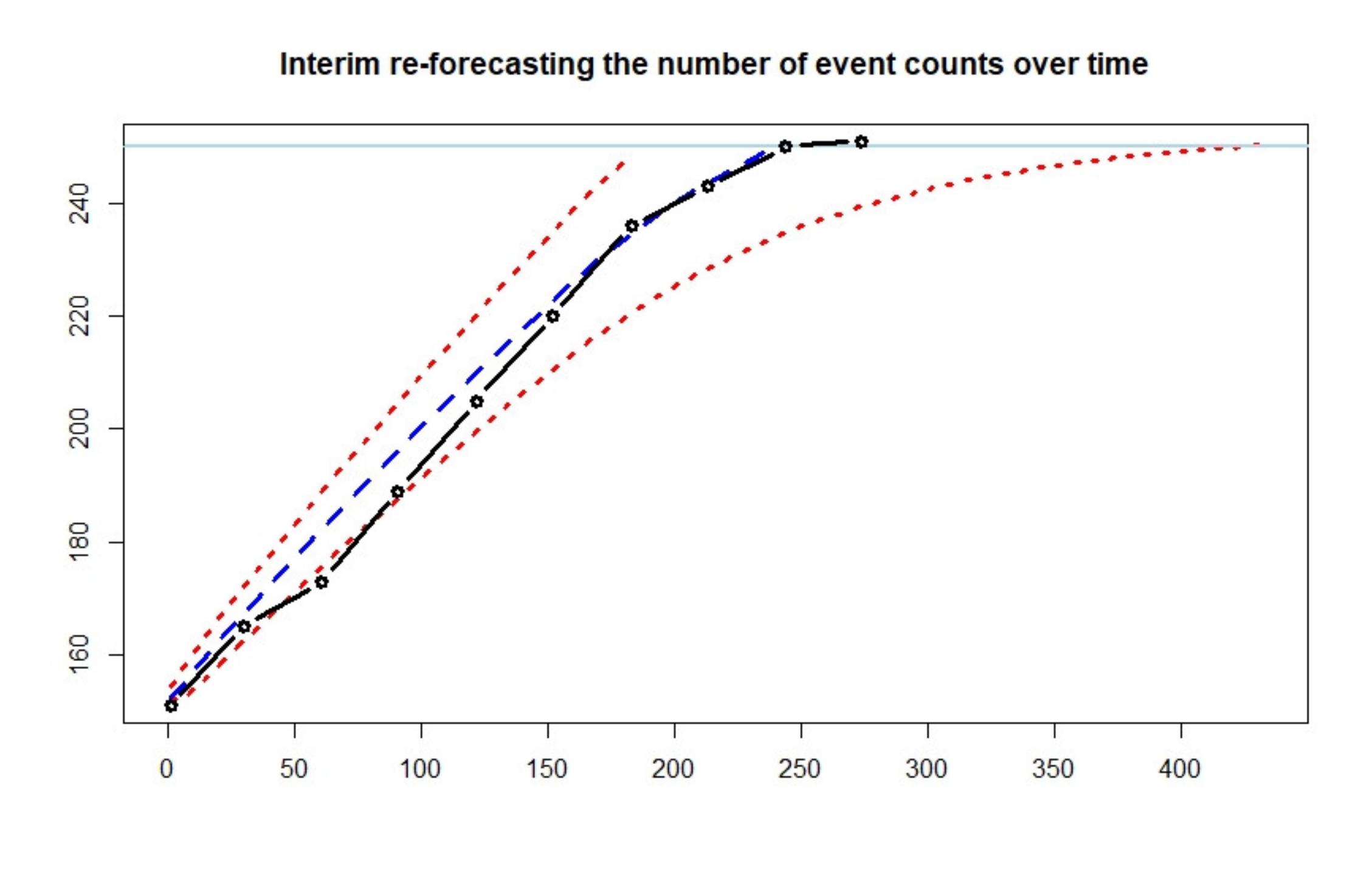}
\caption{Prediction of the remaining number of events against time from cut-off (in days).
Real trajectory of events is shown by black line. Weibull model,
mean prediction and bounds depicted by the blue dashed and red dotted lines.}
\label{Fig-6}
\end{center}
\end{figure}

Fig~\ref{Fig-4} shows very good fit of the predictive area of recruitment where
the real trajectory of the historical recruitment falls into the predictive area.

As one can see from Fig~\ref{Fig-5} and Fig~\ref{Fig-6}, the predictions
for both types of models, exponential and Weibull,
are rather close, with the following estimated parameters for each: \\
Exponential model: $\mu_A = 0.0069$, $\mu_L = 0.00034$
and $r = 0.2566$. \\
Weibull model: $a_A =1.1636$, $b_A = 126.9995$,
$a_L = 0.3177$, $b_L =  2053795.7$ and $r = 0.2996$.

The actual number of days when the target number of patients was reached in this trial was 152 and the actual number of days
when the planned number of events occurred is 244. As seen in the figures, the predictions
were indeed very close to the actuals.

\subsection{Predicting events when recruitment complete}

We also considered the same case study as above, but at a later interim time when recruitment had completed.
Therefore the task here was to predict the future event counts only.

As detailed in the previous section, this study has a target number of events of 250
and patients sample size of 405.
At the interim cut-off time the data
for the study had the following recruitment and event status:
\begin{itemize}
\item 220 Events (i.e. censor\_flag == 1)
\item 163 At Risk (i.e. censor\_flag == 1 and drop\_out\_flag == 0)
\item 22 Drop Outs( i.e. drop\_out\_flag == 1)
\end{itemize}

\subsubsection{Key predictions}
Given the above input and implementing the developed model,
the \textit{EventPrediction} package predicted:

\textit{Events:} Predicted number of days until target number of events
reached, with 90\% bounds, (mean, lower bound, upper bound):\\
exponential model: 84, 60, 118 \\
Weibull model: 83, 59, 116

\subsubsection{Plots and parameter estimates}
Further, the \textit{EventPrediction} package produced the following two key plots,
along with key parameter estimates, for the key stakeholders to consume: \\
1) prediction of the remaining number of events using exponential model \\
2) prediction of the remaining number of events using Weibull model

\begin{figure}[ht]
\begin{center}
\includegraphics[width=11.0cm,height=6cm]{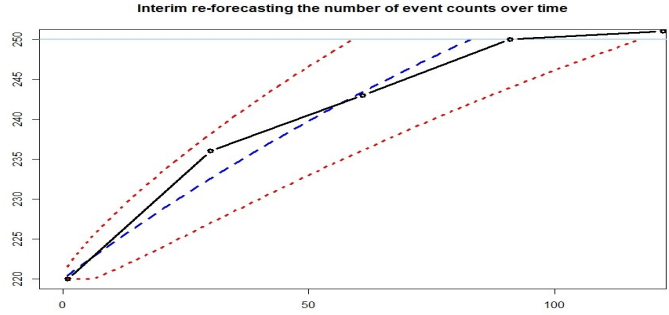}
\caption{Prediction of the remaining number of events against time from cut-off (in days).
Real trajectory of events is shown by black line.
Exponential model, mean prediction and bounds depicted by the blue dashed and red dotted lines.}
\label{Fig-8}
\end{center}
\end{figure}

\begin{figure}[ht]
\begin{center}
\includegraphics[width=11.0cm,height=6cm]{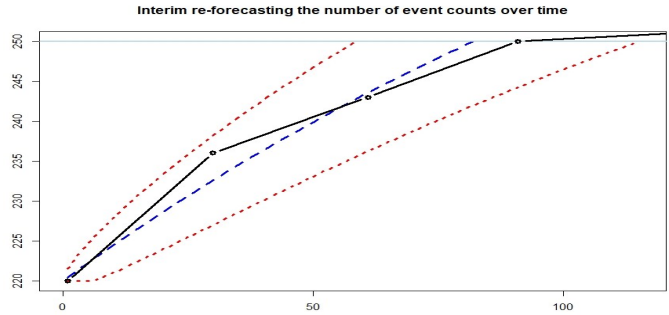}
\caption{Prediction of the remaining number of events against time from cut-off (in days).
Real trajectory of events is shown by black line. Weibull model,
mean prediction and bounds depicted by the blue dashed and red dotted lines.}
\label{Fig-7}
\end{center}
\end{figure}

As one can see from Fig~\ref{Fig-8} and Fig~\ref{Fig-7}, the predictions
for both types of models, exponential and Weibull,
are rather close, with the following estimated parameters for each: \\
Exponential model: $\mu_A = 0.00553$, $\mu_L = 0.00034$
and $r = 0.2128$. \\
Weibull model: $a_A =0.9834$, $b_A = 183.8473$,
$a_L = 0.36351$, $b_L =  349763.5$ and $r = 0.2067$.

The actual number of days when the planned number of events occurred is 91. As seen in the figures, the predictions
were indeed very close to the actuals.

\section{Fitting models to real data}

\subsection{Kaplan-Meier plots}
To assess the model fit, we looked at the Kaplan-Meier (KM) curve for each interim dataset and compared these to
the predicted survival functions for exponential and Weibull distributions. This provides a visualisation
for model fit:
the best fit model for the interim data will be the model which distribution best maps the KM curve.
This is for the occurrence of events $A$ and so will only inform of the
best distribution for modelling events $A$. However, similar curves can be created to test the fit
of dropout distribution.

The survival functions for exponential and Weibull models with cure are calculated
using their respective estimated parameters,
with formulae:

Exponential: $S(x, r, \mu_A) = r + (1 - r) \exp(-\mu_A x)$ \\
Weibull: $S(x, r, \a_A, b_A) = r + (1 - r) \exp(- (\frac{x}{b_A})^{\a_A})$

\begin{figure}[ht]
\begin{center}
\includegraphics[width=11.0cm,height=6cm]{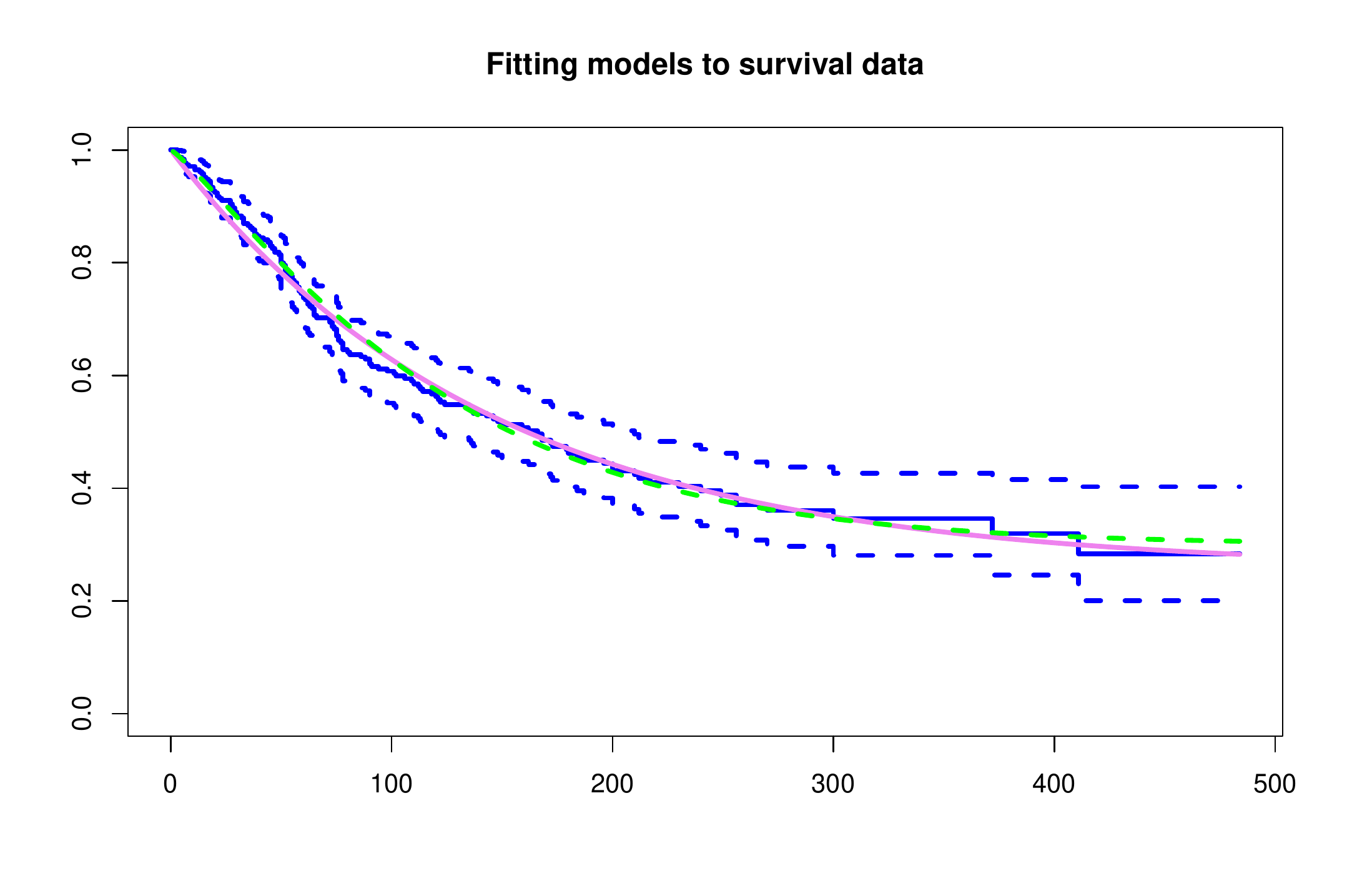}
\caption{Kaplan-Meier plot showing survival function using interim data
with associated confidence intervals (blue lines), alongside survival functions
for exponential (violet solid line) and Weibull (green dashed line) cure models.}
\label{Fig-9}
\end{center}
\end{figure}

In Fig~\ref{Fig-9}, one can see that the predicted survival functions for exponential and Weibull models
are very close and both map the KM curve well.

\subsection{AIC and BIC criteria}

We also used Akaike information criterion (AIC) and Bayesian information criterion (BIC) to test which model best fits the real dataset. These are calculated by:

AIC $= 2k - 2 {\rm LogLik}$

BIC $= k \log(n) - 2 {\rm LogLik}$

where $k$ = number of parameters in the model, $n$ = total number of patients at interim time,
${\rm LogLik}$ = the log-likelihood from parameter estimation.

For the stopped recruitment case:
\begin{center}
\begin{tabular}{l c c}
\textbf{Model} & \quad \textbf{AIC} & \quad \textbf{BIC} \\
Exponential model, no cure & 3333.6&3341.6 \\
Exponential cure model &3319.0&3327.1 \\
Weibull (A) and exponential (L) cure model & 3321.0 & 3333.0 \\
Exponential (A) and Weibull (L) cure model & 3283.1 & 3299.1 \\
Weibull cure model&3285.0&3305.1
\end{tabular}
\end{center}

For the ongoing recruitment case:
\begin{center}
\begin{tabular}{l c c}
\textbf{Model} & \quad \textbf{AIC} & \quad \textbf{BIC} \\
Exponential model, no cure &2222.9 &2230.4 \\
Exponential cure model &2210.9&2218.4 \\
Weibull (A) and exponential (L) cure model & 2209.3 & 2220.6 \\
Exponential (A) and Weibull (L) cure model & 2183.7 & 2198.7  \\
Weibull cure model&2182.1&2200.9
\end{tabular}
\end{center}

The lower the AIC or BIC, the better the model fit. From the two tables above,
one can see that the cure models show an improvement on the ''no cure'' model.
The exponential (A) and Weibull (L) cure model is the best fit for the stopped recruitment dataset.
By small margins, the values of criteria for ongoing recruitment
suggest either the Weibull cure model
or the exponential (A) and Weibull (L) cure model is best suited for the data.

\section*{Conclusions}

We have developed a new analytic approach for the prediction of event counts
in event-driven trials
when recruitment is complete or ongoing.
We use the exponential and Weibull models and account for patient dropout and opportunity of cure.
Not only can we predict the future occurrence of events, but we can also predict
any remaining recruitment, with mean and bounds, using the Poisson-gamma recruitment model.
The developed results can be easily extended to the combined cure models
using the combination of exponential and Weibull distributions for the time to event
and the time to dropout and also to log-normal with cure model.

Using these novel advanced models
and with access to real subject level and centre level data
we are now able to address
key business use cases for a number of key stakeholders
in order to better forecast the operational design of event-driven clinical trials.
Furthermore, by centralising an exposed R Package \textit{EventPrediction},
utilising good software engineering principles, each of our key
stakeholders have access to the package and can obtain plots,
parameter estimates and predictions with bounds,
 without contacting the mathematical modellers nor the R package developer.

We have many opportunities for future improvements
to the mathematical modelling and \textit{EventPrediction} package.
One of the major priorities is to
evaluate the predictions against real clinical
 trial operational data to ensure the existing and
 any future models are as accurate as we found in testing on historical data.
We  are also looking to incorporate other statistical distributions
for modelling time to  both the main event and dropout.

\bibliographystyle{unsrtnat}

\small


\end{document}